\documentclass[12pt]{article}
   \usepackage{amsfonts}
\newcommand{\bc}{\begin{center}}
\newcommand{\ec}{\end{center}}

\def\d{\partial}

\def\pint{\int\limits_{0}^{\infty}{\kern-.81em\hbox{{\sf --}}}~~~}

\makeatletter
\renewcommand{\@makefnmark}{\hbox{\mathsurround=0pt$^{\@thefnmark{})}$}}
\renewcommand{\@makefntext}[1]{\parindent=1em\noindent
\hbox to 1.8em{\hss$^{\@thefnmark{})}$}#1} \makeatother
\begin{document} %
%%%%%%%%%%%%%%%%%%%%%%%%%%%%%%%%%%%%%%%%%%%%%%%%%%%%%%%%%%%%%%%%%%%
%%%%%%%%%%%%%%%%%%%%%%%%%%%%%%%%%%%%%%%%%%%%%%%%%%%%%%%%%%%%%%%%%%%
%%%

\begin{center}
{\Large  The radiation reaction effects in the BMT model of
spinning charge and the radiation polarization phenomenon}
\vspace{10mm} \\
S.L.~Lebedev \footnote{e-mail: sll@chgpu.edu.ru }\\

 \vspace*{10mm}
\begin{minipage}[t]{130mm}
 \bc
 {\small \it Department of Physics, Chuvash State Pedagogical
University, Cheboksary, 428000, Russia} \ec
\end{minipage}
 \vspace{10mm}

 Abstract
 \vspace{4mm}
\\
\begin{minipage}[t]{130mm}
\baselineskip 5.4mm % mark
{\small  The effect of radiation polarization (RP) attended with
the motion of spinning charge in the magnetic field could be
viewed through the classical theory of self-interaction. The
quantum expression for the polarization time follows from
semiclassical relation $T_{QED}\sim
\hbar\,c^{3}/\mu_{B}^{2}\omega_{c}^{3}$, and needs quantum
explanation neither for the orbit nor for the spin motion. In our
approach the polarization emerges as a result of natural selection
in the ensemble of elastically scattered electrons, among which
the group of particles that bear their spins in the 'right'
directions has the smaller probability of radiation.}
%%%%
\end{minipage}
\end{center}
\vspace{0mm} \noindent
%%%
%\newpage
\baselineskip 6mm % mark
% You can change the spacing between lines by changing the above
%parameter.

\def\<{\langle}
\def\>{\rangle}
\def\[{\left[}
\def\]{\right]}
\def\({\left(}
\def\){\right)}
\def\ds{\displaystyle}
\def\ts{\textstyle}
\def\ss{\scriptstyle}
\def\wt{\widetilde}
\def\sh{\mathop{\rm sh}\nolimits}
\def\ch{\mathop{\rm ch}\nolimits}
\def\th{\mathop{\rm th}\nolimits}
\def\diag{\mathop{\rm diag}\nolimits}
\def\ln{\mathop{\rm ln}\nolimits}

%\vspace{5mm}
%\vspace*{6mm}

%\title{The radiation reaction effects in the BMT model of spinning
%charge and the radiation polarization phenomenon}

\bc {\bf 1. Introduction} \ec

The rise in popularity of the classical spin models was stimulated
by the difficulties with high spin wave equations accounting for
the interaction of particle with external EM field \cite{VZ}. The
close relation of the (pseudo)classical models of spinning
particles to the string theory raises a new phase of interest in
this topic \cite{BM,HeT}. The criterion used by different authors
to check the spin degrees of freedom are described correctly, is
the possibility for one to obtain, at least with some
approximations involved, the Bargmann-Michel-Telegdi (BMT) or the
Frenkel-Nyborg equations determining the spin evolution (see e.g.
\cite{BM,Lya}). The reason for this is the universal character of
the BMT equation and its well established experimental
applicability. Here we consider the problem of self-interaction of
the BMT particle and its relation to the RP phenomenon.
\footnote{It is worth noting that the account for radiation
through the local ALD-type equation for spinning charge seems
hardly to have a practical meaning (see the extensive study of
that topic in \cite{RR})}

The effect of preferable polarization emerges when the
relativistic ($\sim\,1\,GeV$ ) electrons execute the motion in
magnetic field during the polarization time \footnote{With some
obvious exceptions we use the system of units with $c=1,\,\, \hbar
=1,\,\, \alpha =e^{2}/4\pi\hbar c ,\,\,
a_{B}=4\pi\hbar^{2}/me^{2}$ and $H_{c}=m^{2}c^{3}/e\hbar$ ($\alpha
,\, \, a_{B}$ and $H_{c}$ being the fine structure constant, the
Bohr radius and critical QED field strength).}
$$
T_{QED}=\frac{8\sqrt{3}}{15}\frac{a_{B}}{c}
\gamma^{-2}\(\frac{H_{c}}{H}\)^{3} \eqno{(1)}
$$
on the laboratory clocks \cite{ST}. RP manifests itself through
the asymmetry of the probability of the spontaneous spin-flip
transitions
$$
w^{\uparrow\downarrow}=
\frac{1}{T_{QED}}\(1+\zeta_{3}\frac{8\sqrt{3}}{15}\)
\eqno{(2)}
$$
w.r.t. the value of the initial polarization $\zeta_{3}=\pm 1$
\cite{T}.  The physical ground for that asymmetry is, of course,
radiation process so that the phenomenon itself could be
considered as a back-reaction effect. The latter one can describe
with the help of the semiclassical elastic scattering probability
\cite{Rit'81}
$$
\exp{\(-\frac{2}{\hbar}\Im\Delta W\)}\,\,<1
\eqno{(3)}
$$
where the classical self-action of the charge \footnote{The
subtraction of UV divergences corresponding to the definition of
the observable mass is implied in (4) \cite{Rit'81,sll'02}.}
$$
\Delta W=\left.\frac12
\int\int\,J_{\mu}(x)\Delta_{c}(x,x')J_{\mu}(x') \,dx\,dx' \right.
\eqno{(4)}
$$
should have a positive imaginary part ($\Im \Delta W\,>0$)
pointing to the presence of radiation. The photon Green function $
\Delta_c(x,x')=i(2\pi)^{-2}/[(x-x')^2+i0] \,$ and the source
$$
J_{\mu}=j_{\mu}+\partial_{\nu}M_{\mu\nu}
\eqno{(5)}
$$
includes the orbit ($j_{\mu}$) and the spin
($\partial_{\nu}M_{\mu\nu}$) contributions.\\
Below we clarify in short technical points of calculations
performed and discuss the results and some differences
from the original considerations in \cite{ST,Ba}.\\

\bc {\bf 2. The mass shift and the internal geometry of the world
lines} \ec

With the help of eqn.(5) the self-action $\Delta W$ could be
decomposed as follows:
$$
\Delta W=\Delta W_{or}+\Delta W_{so}+\Delta W_{ss}\, .
\eqno{(6)}
$$
The orbit part $\Delta W_{or}$ does not contain the spin degrees
of freedom and for the case of constant homogeneous EM field
(which is the matter of interest to us here)was studied in
\cite{Rit'81}. The spin-orbit part
$$
\Delta W_{so}=-\frac{\mu e}{2\pi^2}\int d\tau \int d\tau'
\frac{\varepsilon_{\alpha\beta\gamma\delta}(x-x')_{\alpha}\dot
x_{\beta}(\tau) \dot
x_{\gamma}(\tau')S_{\delta}(\tau')}{[(x-x')^2]^2} \, ,
\eqno{(7)}
$$
was discussed in \cite{sll'02,sll'03}, and
$$
\Delta W_{ss}=\frac{\mu ^{2}}{2}\int d\tau \int d\tau'
\,\,u_{[\beta^{\prime}}S_{\rho]}\,
u^{\prime}_{[\beta}S^{\prime}_{\rho]}\,\d^{\prime}_{\beta^{\prime}}
\d_{\beta} \, \Delta_c(x,x')
 \eqno{(8)}
$$
(unlike $\Delta W_{ss}$ self-action $\Delta W_{so}$ contains no UV
divergences). We use the following notations in (5), (7) and (8):
$$
M_{\alpha\beta}(x)=\int\, d\tau
\mu_{\alpha\beta}(\tau)\,\delta^{(4)}(x-x(\tau)) \eqno{(9)}
$$
is the polarization density; Frenkel polarization tensor
$$
\mu_{\alpha\beta}=i\mu\varepsilon_{\alpha\beta\gamma\delta}\dot
x_{\gamma}(\tau)S_{\delta}(\tau)
\eqno{(10)}
$$
with $\mu=\frac{1}{2}\;g\;\mu_{B}$, and $g$, $\mu_{B}=e/2m$ being
the $g$ factor and Bohr magneton correspondingly. 4-velocities
$u\equiv \dot{x}(\tau)$, $u'\equiv \dot{x}(\tau')$ and spin
4-vectors $S\equiv S(\tau)$, $S'\equiv S(\tau')$ are determined
from Lorentz and BMT equations:
$$
\dot u=\kappa \hat{F}\cdot\, u \,,
 \eqno{(11)}
$$

$$\dot S=\frac{1}{2}\,g\, \kappa\hat F\cdot S+(\frac g2-1)\,\kappa\,
u\,(u\cdot\hat F\cdot S) ,
 \eqno{(12)}
$$
$(\kappa\equiv e/m)$ where the dot from above means derivative
w.r.t. proper time.

For the constant homogeneous background the translational symmetry
entails in
$$
 \Delta W = -\Delta m \,T \, ,
 \eqno{(13)}
$$
with $ \Delta m$ denoting the mass shift (MS) and $T$
corresponding to the proper time interval of the charge's stay in
external field. In application of eqn.(13) it is, generally,
supposed that the formation (proper) time of the non-local $
\Delta m$ is much less than $T$.

The important property of the motion in the constant field is the
'isometry' property of the world lines \cite{Rit'81}:
$$
(x(\tau)-x(\tau'))^{2}=f(\tau-\tau')\, .
\eqno{(14)}
$$
Here the function $f$ is an even function of the proper time
difference $ \Delta \tau=\tau-\tau'$. Given this difference, the
integrands in expressions (7) and (8) preserve their value along
the world line so that these non-local geometrical characteristics
exhibit some kind of 'rigidity' which eventually gives rise to
eqn.(13).

To compute the invariants present as integrands in the
self-actions $ \Delta W_{so}$ and $ \Delta W_{ss}$ one can exploit
Frenet-Serret (FS) formalism adapted to the case of constant
homogeneous EM field in \cite{HSV}. Let ${e^{A}}\, (A=0,...,3)$ be
a FS tetrad with $e^{(0)}\equiv u(\tau)$. For every element of
tetrad the Lorentz equation is valid:
$$
\dot e^{A}=\kappa \hat{F}\cdot\, e^{A}(\tau) \,.
 \eqno{(15)}
$$
Combining the eqn.(15) with the basic equation of FS formalism,
$$
\dot e^{A}=\Phi^{A}_{\;B}\, e^{B}(\tau) \,,
 \eqno{(16)}
$$
one can turn the action of the Lorentzian matrix $\hat{F}$ into
the tetrad basis. Of first importance now is the constancy of
Frenet matrix $\Phi^{A}_{\;B}$. The non-zero elements of
$\Phi^{A}_{\;B}$ are the curvature ($k$), the first ($t_{1}$) and
the second ($t_{2}$) torsions, which have their representations
directly in terms of electric and magnetic fields \cite{HSV}.

\bc {\bf 3. Results} \ec

Below we concentrate on the plane motion in the purely magnetic
field and $g=2$ $-$ those conditions are simplest one to make
possible the comparison with standard QED \cite{ST} and
semiclassical QED \cite{Ba} approaches. For that case
$k=\kappa\,H\,v_{\bot}\gamma_{\bot}=v_{\bot}\,t_{1}$ and
$t_{2}=0$, so that the MS $\Delta m_{so}$ and $\Delta m_{ss}$
corresponding to the self-actions (7) and (8) can be transformed
into forms:
$$
\Delta m_{so}=-i\frac{\mu e}{2\pi^2}\,S_3\,\omega_c^2\,
f_{so}(v_{\perp})\, , \eqno{(17)}
$$
$$
 \Delta m_{ss}=\left\{
\begin{array}{cc}
k_{0}+k_{13}-k_{13}\;\zeta_{3}^{2}+(k_{12}-k_{13})\; \zeta_{\bot
v}^{2} \, ,\\ \  \\
k_{0}+k_{12}-k_{12}\;\zeta_{3}^{2}-(k_{12}-k_{13})\; \zeta_{v}^{2}
\, .
\end{array} \right.
\eqno{(18)}
$$
The upper and lower representations in eqn.(18) are equivalent
since the spin vector $\vec{\zeta}$ in the rest frame of the
particle satisfies the relation
$$
\vec{\zeta}^{2}=\zeta_{3}^{2}+\zeta_{v}^{2}+\zeta_{\bot v}^{2}=1
\,.
\eqno{(19)}
$$
Note that $S_{3}=\zeta_{3}$, $S_{v}=\gamma_{\bot}\zeta_{v}$, and
$S_{\bot v}=\zeta_{\bot v}$ are the (conserved) spin components
parallel to the field $\mathbf{H}$, parallel to the velocity
${\mathbf{v}}(={\mathbf{v_{\bot}}})$ and perpendicular to
${\mathbf{v}}$ correspondingly. The following notations were used
in formulas (17) and (18):
$$
f_{so}=\frac{v^2_{\perp}}{\gamma_{\perp}}\int^{\infty}_{0}\,
\frac{\sin^2{w}-w\sin{w}\cos{w}}{(v^2_{\perp}\sin^2{w}-w^2)^2}\,dw
\, ,
\eqno{(20)}
$$
$$
k_{12}=\frac{-i\mu^2\omega_{c}^{3}}{4\pi^{2}\gamma_{\perp}^{4}}
\int^{\infty}_{0}\,
\[\frac{-w^{2}\sin^2{w}}{(w^{2}-v^2_{\perp}\sin^2{w})^3}+
\frac{\gamma_{\perp}^{6}}{w^{2}}\]\,dw \, , \eqno{(21)}
$$
$$
k_{12}-k_{13}=\frac{-i\mu^2\omega_{c}^{3}v^{2}_{\perp}}
{4\pi^{2}\gamma_{\perp}^{2}} \int^{\infty}_{0}\,
\frac{(w\cos{w}-\sin{w})^{2}}{(w^{2}-v^2_{\perp}\sin^2{w})^3} \,dw
\,  \eqno{(22)}
$$
with $\omega_{c}=eH/m$ and $ \gamma_{\bot}$ being the Lorentz
factor. The term $k_{0}+k_{13}$ ($k_{0}+k_{12}$) in the upper
(lower) part of the eqn.(18) do not depend on the spin direction
and is not of importance in explaining RP. As the numerical
investigation shows, the functions $k_{12}$ and $k_{13}$ are
rather close each other. Note also, that $i(k_{12}-k_{13})$ is
positive in no dependence on the energy of the particle as well as
the functions $ik_{12}$ and $ik_{13}$ itself.

\bc {\bf 4. Discussion} \ec

The probability of not-emitting the photons is decreasing with the
proper time according to general law (see (3) and (13)):
$$
\exp{(\Im\Delta m \cdot T)}\,,\qquad\qquad\Im\Delta m < 0\, .
 \eqno{(23)}
$$
Accounting for the positivity of the integrals in the r.h.s. of
expressions (21), (22) one can guess from the eqn.(18) that
particles with $\zeta_{v}\neq 0$ would have a better chance to
preserve their state whereas the particles with $\zeta_{\bot
v}\neq 0$ such a possibility should lose just with the same rate.

Supposing the relativistic energies for electrons we find for the
spin-dependent part of the total MS $\Delta m =$ $\Delta
m_{or}+\Delta m_{so}+\Delta m_{ss}$ the following sum

$$
-i\frac{1}{4\sqrt{3}\,a_{B}}\,\chi^{2}\zeta_{3}+
i\frac{15}{64\sqrt{3}\,a_{B}}\,\chi^{3}\zeta_{3}^{2}+
i\frac{1}{64\sqrt{3}\,a_{B}}\,\chi^{3}\zeta_{v}^{2}\, ,
 \eqno{(24)}
$$
where the first term comes from $\Delta m_{so}$ (it would have an
opposite sign for positrons \cite{sll'03}) and
$\chi=\gamma_{\bot}H/H_{c}$. With the notation
$$
\lambda=-\frac{2}{\hbar}\Im\Delta m c^{2} \eqno{(25)}
$$
we arrive at the spin contribution to the decay rate in the form:
$$
\lambda_{spin}=
\frac{c}{a_{B}}\chi\[\frac{1}{2\sqrt{3}}\,\chi\zeta_{3}-
\frac{15}{32\sqrt{3}}\,\chi^{2}\zeta_{3}^{2}-
\frac{1}{32\sqrt{3}}\,\chi^{2}\zeta_{v}^{2}\] \, . \eqno{(26)}
$$
$\lambda_{spin}$ would be negative for $\zeta_{3}<0$. This, of
course, has no effect  on the positivity of the total 'decay rate'
$\lambda$ since $\chi\ll 1$.

 Being the probability of radiation per unit
proper time, $\lambda$ in (25) corresponds to either change of
particle's state of motion- not only to the spin-flip transitions.
The negative $\zeta_{3}$ slightly reduces this probability, as
well as two last terms in (26) do - in no dependence on the signs
of $\zeta_{3}$ and $\zeta_{v}$. Note, that according to eqns.(18)
and (26) spin-spin interaction itself does not give the preferable
polarization for elastically scattering particles ('down' for
electrons and 'up' for positrons). The RP effect emerges in
conjunction of the spin-orbit and spin-spin interactions. The
characteristic laboratory times extracted from (26) are
$$
T^{(1)}_{ss}=\frac{32\sqrt{3}}{15}\,\frac{a_{B}}{c}\chi^{-3}
\gamma_{\bot}\, ,\qquad T^{(2)}_{ss}=15\,T^{(1)}_{ss}\, .
\eqno{(27)}
$$
The presence of the last term in (26) corresponds to the
incomplete polarization degree among the elastically scattered
electrons estimated as $\sim 15/16= 0.938$ (compare it with the
dynamic value $0.924$ of the polarization degree in QED). The
relation $T^{(1)}_{ss}=4\,T_{QED}$ one finds from (1) and (27),
should be addressed to the lack of the direct correspondence
between $\lambda$ and $w^{\uparrow\downarrow}$ in (2). The
classical model of spin relaxation proposed in \cite{JDJ,BTB,T}
gives for the polarization time $T_{QED}$ the wanted expression up
to the factor of order unity (not four) and for the polarization
degree $100\%$. So, in what concerns classical consideration,
 the 'nice' "4" and $0<T^{(2)}_{ss}<\infty$
(see (27)) are the main variation of our results from those of
\cite{JDJ,BTB,T}.

The author would like to acknowledge useful discussions with
V.I.~Ritus and financial support from  grant SS 1578.2003.2.
\vspace{-0.8 cm}
\newpage
\bc {\bf 5. References} \ec

\end{document}